\documentclass[cameraready]{Interspeech}

\title{Position-Aware Target Speaker Extraction for Long-Form Multi-Party Conversations: A Diarization-Free Framework for ASR}

\author[affiliation={1}, orcid=0009-0003-9852-6560]{Yichi}{Wang}
\author[affiliation={1}, orcid=0009-0004-7125-2296]{Junzhe}{Chen}
\author[affiliation={1}, orcid=0009-0008-0102-3550]{Wangjin}{Zhou}
\author[affiliation={1}, orcid=0000-0002-2686-2296]{Tatsuya}{Kawahara}

\address{
    $^1$ Graduate School of Informatics, Kyoto University, Kyoto, Japan
}

\email{\{yiwang, juchen, zhou, kawahara\}@sap.ist.i.kyoto-u.ac.jp}


\usepackage{comment}
\usepackage{amsmath}
\usepackage{amsfonts}
\usepackage{amssymb}
\usepackage{booktabs}
\usepackage{array}
\usepackage{diagbox}
\usepackage{makecell}
\usepackage{graphicx} 
\usepackage{multirow}
\usepackage{makecell}
\usepackage{tabularx}
\usepackage{array}
\usepackage{pifont}
\usepackage{hyperref}

\newcolumntype{L}{>{\raggedright\arraybackslash}X}
\newcolumntype{C}{>{\centering\arraybackslash}X}

\newcommand{\asep}{\!/} 
\newcommand{\angcell}[4]{\mbox{#1\asep#2\asep#3\asep#4}}

\begin{document}

\maketitle

\begin{abstract}
In long-form multi-party conversations, highly imbalanced speaker activity and frequent overlap make it difficult to identify “who spoke when and what”. 
Sliding-window continuous speech separation (CSS) mitigates sparse supervision, but often suffers from cross-window speaker inconsistency and residual crosstalk, which in practice requires diarization for reliable speaker attribution. 
Motivated by the stability of speakers' directions of arrival (DOAs) in meetings, we propose \textbf{PATSE}, a multi-channel \textbf{P}osition-\textbf{A}ware \textbf{T}arget \textbf{S}peaker \textbf{E}xtraction front-end that uses DOA as a spatial prior to directly extract the speech of each target speaker.
PATSE combines a DOA-guided spatial encoder and conditioner to generate speaker-attributed streams, from which speaker activity can be inferred via simple post-processing (e.g., VAD) without explicit diarization.
Experiments on both replayed and real conversations show consistent ASR gains outperforming CSS and diarization-based pipelines.

\end{abstract}

\section{Introduction}


In multi-party conversations such as meetings and discussions, the fundamental problem is to identify “who spoke when and what”~\cite{cornell2024one,shi2026train,shi2023casa}. 
Real-world recordings are typically long and continuous, where spontaneous conversations results in highly imbalanced speaker activity, and back-channel responses and brief interruptions cause frequent speech overlap.
Since most automatic speech recognition (ASR) systems are optimized for short single-speaker utterances, a widely-used strategy is to introduce a speaker-aware separation front-end to enable speaker-attributed transcription.


Most separation systems allocate one output per speaker. 
For long recordings, however, many speakers remain inactive for long periods, yielding temporally sparse supervision and unstable training~\cite{niu2025dcf}. 
Continuous speech separation (CSS)~\cite{chen2020continuous} mitigates this issue by applying separation in sliding windows where only a subset of speakers are active. 
However, consistent speaker identity across discontinuous windows remains difficult: similarity-based linking is local, while separation alone does not enforce long-range identity consistency, so speaker diarization (SD) is typically required for reliable assignment. 
In addition, CSS is prone to producing crosstalk in single-speaker regions~\cite{raj2021integration, taherian2023multi}, causing insertion errors in downstream ASR.

The interaction between separation and diarization remains an open problem~\cite{yoshioka2019low, von2021graph}. 
Some approaches perform diarization prior to separation~\cite{delcroix2021speaker}, as in guided source separation~\cite{raj2022gpu,medennikov2020stc}. 
However, accurate estimation of speaker temporal boundaries in overlapping speech remains challenging~\cite{fujita2019end}. 

Since ideal separation naturally captures speaker activity, joint modeling has been explored to overcome cascaded limitations~\cite{taherian2024multi}. 
As a form of joint modeling, 
we formulate the front-end as target speaker extraction (TSE).
TSE extracts a specified speaker’s speech conditioned on prior target information~\cite{briegleb2023exploiting,elminshawi2023beamformer,delcroix2019end}, such as speaker embeddings or spatial clues. 
Applying TSE independently to each target yields speaker-attributed streams, which (i) preserves consistent speaker tracking without explicit diarization, and (ii) avoids inter-output crosstalk by producing a single output per target speaker.

Nevertheless, speaker embedding-based TSE has two inherent drawbacks: reliance on enrollment audio~\cite{delcroix2021speaker,wang2018voicefilter}, which is often unavailable in practice, and mismatch between fixed reference embeddings and time-varying speaker characteristics.
Although embeddings can be derived from diarization-based single-speaker segments~\cite{boeddeker2024ts}, noise and interference in these segments may distort the resulting embeddings.
In contrast, spatial cues such as direction of arrival (DOA) provide a more stable and explicit prior~\cite{zhang2025doa}. 
In typical meetings, speakers remain largely stationary, yielding relatively consistent DOAs that can be obtained via microphone arrays or cameras. 
These observations motivate multi-channel position-aware TSE as a practical front-end design for long-form multi-party ASR, extending prior DOA-based TSE studies~\cite{elminshawi2023beamformer, wang2024study, wang2025leveraging} beyond simulated short utterances.

Following are the main contributions of this paper:

\begin{itemize}

    \item We propose PATSE, a multi-channel position-aware TSE front-end for multi-party ASR, featuring a DOA-guided spatial encoder and conditioning module that injects target-specific spatial features into the separation backbone. PATSE produces speaker-attributed streams, from which speaker activity can be inferred through simple post-processing (e.g., voice activity detection, VAD) without explicit diarization.
    
    \item We build and release LibriReplay-DOA\footnote{https://huggingface.co/datasets/real-recordings/LibriReplay-DOA}, a real-room playback dataset with ground-truth DOA annotations to benchmark DOA-based methods. Experiments on LibriReplay-DOA and the real-world conversational TEIDAN dataset~\cite{elmers2025triadic} demonstrate consistent improvements in downstream ASR. Detailed case studies are available on our demo page\footnote{https://exp-demos.github.io/PATSE-audio-demo}.
\end{itemize}

\begin{figure*}[t]
  \centering
  \includegraphics[width=0.75\linewidth]{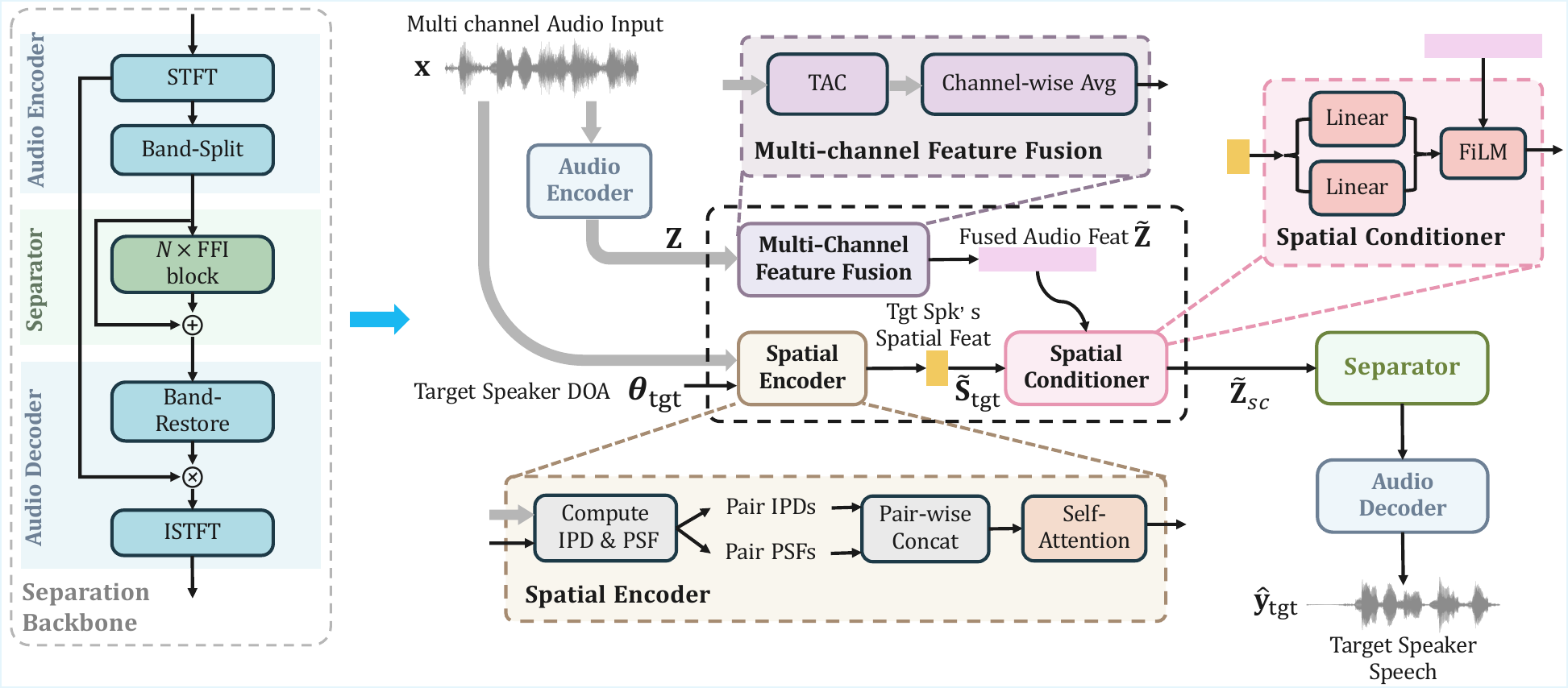}
  \caption{Overall Architecture of the PATSE Framework.}
  \label{fig:framework}
\end{figure*}

\section{Proposed Method}

\subsection{Architecture}
As shown in Figure~\ref{fig:framework}, PATSE treats DOA as an explicit cue for “who” and conditions the separation backbone on this spatial prior.
Let $\mathbf{x} = \{\mathbf{x}_m\}_{m=1}^M$ denote the multi-channel audio signal captured by $M$ microphones, where $m$ indexes the channels.
The azimuth angle $\theta_{\text{tgt}}$ denotes the DOA of the target speaker.

PATSE performs DOA-conditioned extraction via the separation backbone (audio encoder with multi-channel feature fusion (MCFF), separator, and audio decoder), along with a spatial encoder and spatial conditioner:
{
\setlength{\abovedisplayskip}{4pt}
\setlength{\belowdisplayskip}{4pt}
\begin{gather}
\tilde{\mathbf{Z}} = \operatorname{MCFF}(\operatorname{AudioEnc}(\mathbf{x})), \quad
\tilde{\mathbf{S}}_{\text{tgt}} = \operatorname{SpatialEnc}(\mathbf{x}, \theta_{\text{tgt}}), \notag \\
\tilde{\mathbf{Z}}_{sc} =
\operatorname{SpatialConditioner}(\tilde{\mathbf{Z}}, \tilde{\mathbf{S}}_{\text{tgt}}), \tag{1} \\
\hat{\mathbf{y}}_{\text{tgt}} =
\operatorname{AudioDec}(\operatorname{Separator}(\tilde{\mathbf{Z}}_{sc})), \notag
\end{gather}
\setcounter{equation}{1}
\par\raggedright\noindent
}%
where $\tilde{\mathbf{Z}}$ denotes the encoded audio features,
$\tilde{\mathbf{S}}_{\text{tgt}}$ the target-specific spatial features, 
$\tilde{\mathbf{Z}}_{sc}$ the spatially conditioned features, 
and $\hat{\mathbf{y}}_{\text{tgt}}$ the reconstructed waveform of the target speaker.
\subsubsection{Separation Backbone}
In this work, we adopt TIGER~\cite{xu2024tiger} as the separation backbone, an encoder–separator–decoder architecture originally designed for single-channel input.

The shared \textbf{audio encoder} operates on each channel independently. 
For channel $m$, the input waveform $\mathbf{x}_m$ is first transformed into the short-time Fourier transform (STFT) representation $\mathbf{X}_m \in \mathbb{R}^{F \times T}$, where $F$ and $T$ denote the numbers of frequency bins and time frames, respectively. 
The STFT $\mathbf{X}_m$ is then passed through the band-split module to obtain sub-band features $\mathbf{Z}_m \in \mathbb{R}^{N \times K \times T}$, where $K$ is the number of sub-bands and $N$ is the feature dimension per sub-band.
Collecting all channels yields multi-channel audio features
$\mathbf{Z} = \{\mathbf{Z}_m\}_{m=1}^{M}$.

The \textbf{separator} then stacks frequency--frame interleaved (FFI) blocks with residual connections.

The \textbf{audio decoder} finally reconstructs the time-domain waveform $\hat{\mathbf{y}}_{\text{tgt}}$ via band-restore and inverse STFT (iSTFT).

\subsubsection{Multi-Channel Feature Fusion (MCFF)}
To adapt multi-channel features to a monaural separator, 
MCFF fuses $\mathbf{Z}$ into a unified single-stream representation $\tilde{\mathbf{Z}}$.

Specifically, MCFF first adopts a transform--average--concatenate (TAC)~\cite{luo2020end} strategy, where a learnable transform $\mathcal{T}(\cdot)$ is applied to each channel and averaged to form a global descriptor $\mathbf{h}$:
\begin{equation}
\mathbf{h} = \frac{1}{M}\sum_{m=1}^{M}\mathcal{T}(\mathbf{Z}_m).
\end{equation}
The per-channel features $\mathbf{Z}_m$ are concatenated with the global descriptor $\mathbf{h}$ and fused through a mapping $\mathcal{F}(\cdot)$:
\begin{equation}
\mathbf{Z}_m' = \mathcal{F}(\mathrm{Concat}(\mathbf{Z}_m,\,\mathbf{h})) + \mathbf{Z}_m,
\end{equation}
where $\mathbf{Z}_m'$ denotes the globally fused feature of channel $m$.
The final single-stream encoded audio features $\tilde{\mathbf{Z}}$ are obtained by channel averaging:
\begin{equation}
\tilde{\mathbf{Z}} = \frac{1}{M}\sum_{m=1}^{M}\mathbf{Z}_m', 
\qquad \tilde{\mathbf{Z}} \in \mathbb{R}^{N \times K \times T}.
\end{equation}

\subsubsection{Spatial Encoder}

\noindent\textbf{Interaural Phase Difference (IPD): } 
For each microphone pair $p=(i,j)$ ($1 \le i < j \le M$), 
the IPD is defined as:
\begin{equation}
\mathrm{IPD}^{(p)}(t,f)
=
\angle X_i(t,f)
-
\angle X_j(t,f).
\end{equation}
Here $t=1,\dots,T$ and $f=1,\dots,F$ denote the time-frame and frequency-bin indices, respectively.

To avoid phase wrapping ambiguity~\cite{wang2018combining}, it is encoded as:
\begin{equation}
\mathbf{\phi}^{(p)}(t,f)
=
\big[
\cos(\mathrm{IPD}^{(p)}(t,f)),
\sin(\mathrm{IPD}^{(p)}(t,f))
\big].
\end{equation}

\noindent\textbf{Theoretical Phase Difference (TPD):}
TPD denotes the theoretical phase difference induced by a source located at DOA $\theta_{\text{tgt}}$ under the far-field assumption, defined as~\cite{wang2018combining}:
\begin{equation}
\mathrm{TPD}^{(p)}_{\text{tgt}}(f)
=
\frac{2\pi f_s f}{2(F-1)c}\,
cos\theta_{\text{tgt}}d_{i,j},
\end{equation}
where $f_s$ denotes the sampling rate of the waveform, $d_{i,j}$ the distance between microphone $i$ and $j$, and $c$ the speed of sound.

\noindent\textbf{Phase Similarity Feature (PSF): } 
A phase similarity feature between TPD and IPD is defined to quantify the dominance of the target speaker at each time–frequency $(t,f)$ bin~\cite{wang2024study}:
\begin{equation}
\mathrm{PSF}^{(p)}_{\text{tgt}}(t,f)
=
\begin{aligned}
\big[
&\sin(\mathrm{TPD}^{(p)}_{\text{tgt}}(f) - \mathrm{IPD}^{(p)}(t,f)), \\
&\cos(\mathrm{TPD}^{(p)}_{\text{tgt}}(f) - \mathrm{IPD}^{(p)}(t,f))
\big].
\end{aligned}
\end{equation}
It is concatenated with $\mathbf{\phi}^{(p)}(t,f)$ to form the spatial feature:  
\begin{equation}
\mathbf{S}^{(p)}_{\text{tgt}}(t,f)
=
\big[
\mathbf{\phi}^{(p)}(t,f),
\mathrm{PSF}^{(p)}_{\text{tgt}}(t,f)
\big].
\end{equation}
Stacking all $P=\frac{M(M-1)}{2}$ pairs yields $\mathbf{S}_{\text{tgt}}=\{\mathbf{S}^{(p)}_{\text{tgt}}\}_{p=1}^P$.

Following the band-split strategy, stacked self-attention is applied within each band for refinement, and the outputs are aggregated to obtain the final target-specific spatial features $\tilde{\mathbf{S}}_{\text{tgt}}$.

\subsubsection{Spatial Conditioner}
We adopt feature-wise linear modulation (FiLM)~\cite{perez2018film}, 
where a linear generator $\mathcal{L}$ produces modulation parameters 
$\gamma$ and $\beta$ from $\tilde{\mathbf{S}}_{\text{tgt}}$, 
which are used to modulate $\tilde{\mathbf{Z}}$ to obtain $\tilde{\mathbf{Z}}_{\mathrm{sc}}$:
\vspace{-3.5pt}
\begin{equation}
(\gamma,\beta)=\mathcal{L}(\tilde{\mathbf{S}}_{\text{tgt}}), \qquad \gamma,\beta\in\mathbb{R}^{N\times K\times T},
\end{equation}
\begin{equation}
\tilde{\mathbf{Z}}_{\mathrm{sc}}=\gamma\odot \tilde{\mathbf{Z}}+\beta,
\end{equation}
\vspace{-3.5pt}
where $\odot$ denotes element-wise multiplication.

\subsection{Training Objective}
\label{object}

To better supervise the model under different target activity states (speech vs.\ silence), we adopt the activity-aware loss in~\cite{pan2022usev}.
Based on the reference waveform $\mathbf{y}_{\text{tgt}}$, it is segmented into silent and non-silent regions, and the same segmentation is applied to the estimated waveform $\hat{\mathbf{y}}_{\text{tgt}}$, yielding aligned pairs 
$(\hat{\mathbf{y}}_{\text{tgt}}^{\text{sil}}, \mathbf{y}_{\text{tgt}}^{\text{sil}})$ 
and 
$(\hat{\mathbf{y}}_{\text{tgt}}^{\text{n-sil}}, \mathbf{y}_{\text{tgt}}^{\text{n-sil}})$.

\vspace{0.2\baselineskip}
\noindent\textbf{Silent regions:} 
The residual log-energy loss $L_E$ is defined as:
\begin{equation}
L_E = 10 \log_{10}( \lVert \hat{\mathbf{y}}_{\text{tgt}}^{\text{sil}} \rVert^2 + \epsilon ),
\end{equation}
where $\epsilon$ is a small constant for numerical stability.

\vspace{0.1\baselineskip}

\noindent\textbf{Non-silent regions:} 
The signal-to-noise ratio (SNR) loss $L_S$ is defined as:
\vspace{-3.5pt}
\begin{equation}
L_S = -10 \log_{10}(
\frac{\lVert \mathbf{y}_{\text{tgt}}^{\text{n-sil}} \rVert^2}
{\lVert \hat{\mathbf{y}}_{\text{tgt}}^{\text{n-sil}} - \mathbf{y}_{\text{tgt}}^{\text{n-sil}} \rVert^2 + \epsilon}
).
\end{equation}

\noindent{The \textbf{overall loss} $L$ is defined as:}
\vspace{-4pt}
\begin{equation}
L = \alpha L_E + L_S,
\label{eq:loss}
\end{equation}
\vspace{-4pt}
where $\alpha$ is a scaling factor balancing the $L_E$ and $L_S$.












\section{Experiments}

\subsection{Datasets}

\begin{figure}[t]
  \centering
  \includegraphics[width=0.8\linewidth]{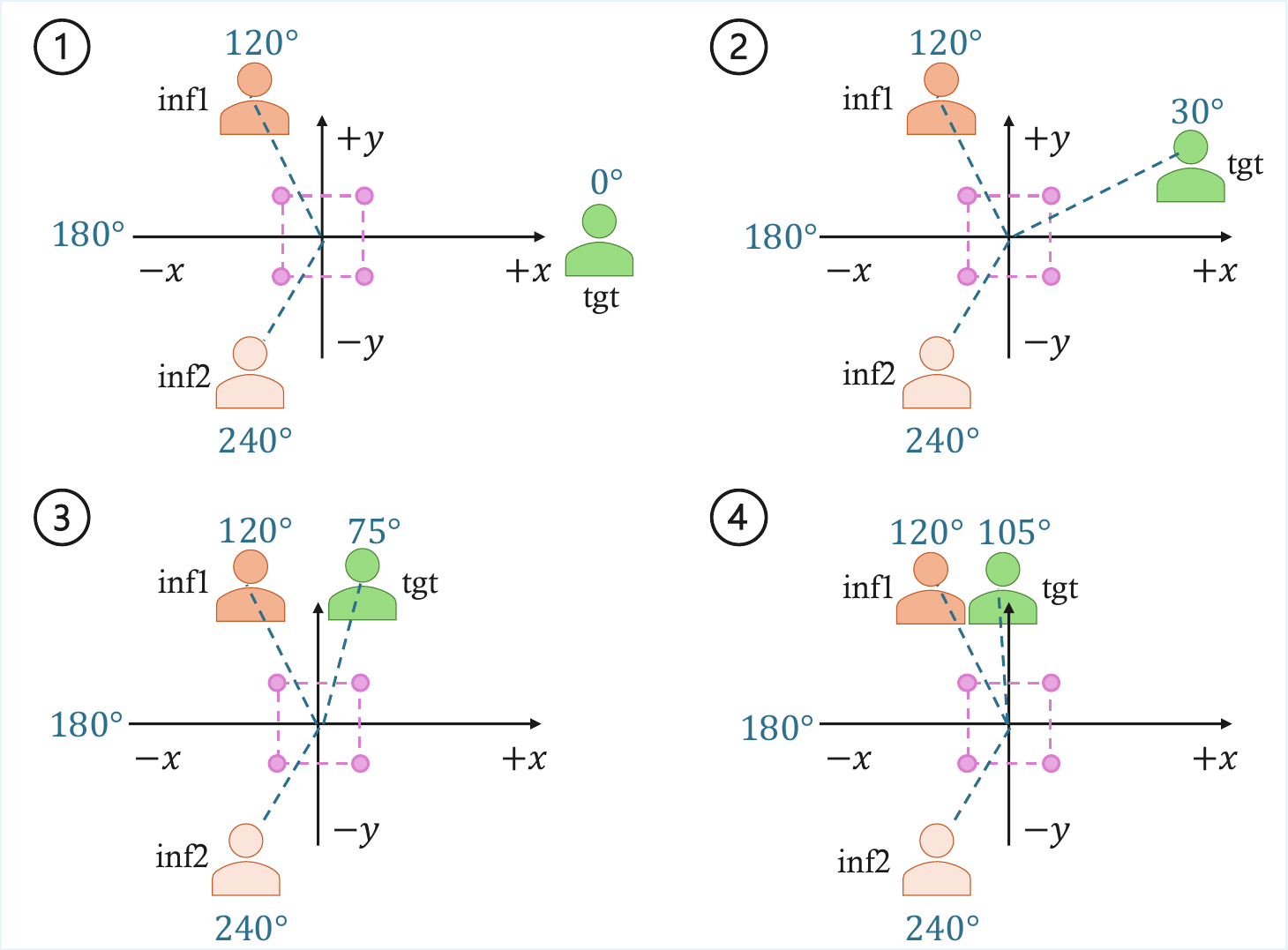}
  \caption{Four Spatial Configurations for LibriReplay-DOA.}
  \label{fig:data}
\end{figure}

\noindent\textbf{LibriReplay-DOA Dataset.}\label{data_gen} Most speech separation benchmarks rely on simulated Room Impulse Responses (RIRs) that may not fully reflect real-world conditions, while real-world corpora often lack ground-truth DOA labels.
To evaluate DOA-based methods reliably, we construct LibriReplay-DOA, a real-room playback dataset with ground-truth DOA annotations.



Multi-party conversations were simulated by concatenating LibriSpeech utterances~\cite{panayotov2015librispeech} with inserted silences for each speaker and replaying them through loudspeakers in a real room.
A total of 114 playback sessions (about 1 minute each) were created, each involving two or three speakers, with overlap ratios of 0--25\% (30), 25--50\% (28), 50--75\% (30), and 75--100\% (26).
Recordings were captured using a four-channel circular array (ReSpeaker Mic Array V2.0, radius 0.032\,m), and DOA labels were given by the physical loudspeaker directions.

To evaluate spatial robustness, four angular configurations were designed by varying the target–interferer angle (15°, 45°, 90°, 120°) while fixing the interferer–interferer angle at 120°, as shown in Figure~\ref{fig:data} (in two-speaker sessions, inf2 remained silent throughout).
Each session was recorded under all four angular configurations, yielding 456 recordings ($\approx$7 hours in total).
\vspace{0.4\baselineskip}

\noindent\textbf{TEIDAN Dataset.} TEIDAN is a real-world triadic dialogue corpus~\cite{elmers2025triadic}, where three participants sit around a circular table (about 120$^\circ$ apart) with a four-channel circular microphone array (radius 0.032\,m) placed at the center.
Experiments are conducted on the English subset (Sessions 01–05), which comprises approximately 2.5 hours of spontaneous dialogue, with each recording lasting about 10 minutes and overlapping speech accounting for 31.98\% of the total duration on average.
\vspace{0.4\baselineskip}


\noindent\textbf{Training Data.}\label{sec:train_data} 
Training (50 h) and validation (6 h) data were generated following the same dialogue simulation procedure as in LibriReplay-DOA. 
For spatialization, RIRs were simulated using the gpuRIR toolbox~\cite{diaz2021gpurir}. 
Rooms were randomly sampled with sizes ranging from $3\times3\times2.4$\,m to $10\times8\times4$\,m and RT60 values between 0.2–0.8\,s. 
A fixed four-channel circular array (radius 0.032\,m) was placed with random position and orientation. 
Speakers were positioned around a virtual table with inter-speaker distances ($\ge 0.3\,\mathrm{m}$), and relative speaker SNRs were sampled from 0–5\,dB. 
Noise samples from the DNS corpus~\cite{reddy2020interspeech} were added as additional point sources, and the mixture SNR was uniformly sampled from 0–20\,dB.








\subsection{Experimental Setup}
\label{sec:exp_setup}

{
\begin{table}[t]
\centering
\caption{Comparison of different front-end methods.}
\label{table:compare}

\fontsize{8pt}{9.5pt}\selectfont
\setlength{\tabcolsep}{3pt}

\begin{tabularx}{\columnwidth}{l c X X X c}
\toprule
\textbf{Method} & \textbf{DOA} & \textbf{Chunk} & \textbf{Stitch} & \textbf{Spk. ID} & \textbf{VAD} \\
\midrule

DSB + Gate & \checkmark & None & -- & -- & Silero \\
FastMNMF & $\times$ & None & -- & \textbf{Oracle} & Silero \\
Sortformer + GSS & $\times$ & SD & -- & SD & -- \\
CSS (FasNet-TAC) & $\times$ & Uniform & MSE & \textbf{Oracle} & -- \\
CSS (TIGER) & $\times$ & Uniform & MSE & \textbf{Oracle} & -- \\
PATSE & \checkmark & Uniform & Direct & -- & Silero \\

\bottomrule
\end{tabularx}
\end{table}
}

{
\begin{table*}[t]
\centering
\caption{WER (\%)$\downarrow$ on the LibriReplay-DOA dataset. 
Results are reported under four target–interferer angles (15°, 45°, 90°, 120°). 
Each entry in the form xx/xx/xx/xx corresponds to overlap ratios of 
0–25\%, 25–50\%, 50–75\%, and 75–100\%, respectively.}
\label{table:LibriReplay}
\fontsize{8pt}{10pt}\selectfont
\setlength{\tabcolsep}{4.8pt}
\renewcommand{\arraystretch}{1.05}

\begin{tabular}{@{}%
l
c
c
c
c
c
c
@{}}
\toprule

\multirow{2}{*}{\textbf{Method}} &
\multirow{2}{*}{\makecell{\textbf{Training}\\\textbf{Strategy}}} &
\multicolumn{4}{c}{\textbf{Speaker angle}} &
\multirow{2}{*}{\textbf{Overall}}
\\
\cmidrule(lr){3-6}
& &
\textbf{15$^\circ$} &
\textbf{45$^\circ$} &
\textbf{90$^\circ$} &
\textbf{120$^\circ$} &
\\

\midrule

DSB + Gate
& NT 
& \angcell{37.1}{ --}{ --}{ --}
& \angcell{33.3}{ --}{ --}{ --}
& \angcell{28.3}{ --}{ --}{ --}
& \angcell{27.3}{ --}{ --}{ --}
& (31.5)
\\

FastMNMF
& NT 
& \angcell{31.3}{ 39.4}{ 40.9}{ 34.7}
& \angcell{14.3}{ 16.5}{ 15.0}{\textbf{ 12.9}}
& \angcell{10.4}{ 22.2}{ 20.1}{ 19.4}
& \angcell{12.7}{ 14.7}{ 15.8}{ \textbf{14.1}}
& 21.1
\\

Sortformer + GSS
& NT 
& \angcell{28.3}{ 28.9}{ 45.5}{ 56.5}
& \angcell{25.1}{ 23.1}{ 42.4}{ 49.1}
& \angcell{22.3}{ 24.4}{ 44.2}{ 51.2}
& \angcell{22.0}{ 24.7}{ 41.5}{ 48.6}
& 38.4
\\

\midrule

CSS (FasNet-TAC)
& Scratch 
& \angcell{46.0}{ 49.9}{ 60.9}{ 61.6}
& \angcell{42.6}{ 39.7}{ 56.2}{ 57.0}
& \angcell{34.9}{ 40.4}{ 56.1}{ 57.3}
& \angcell{30.2}{ 34.2}{ 48.6}{ 48.9}
& 49.5
\\

\midrule




\multirow{2}{*}{CSS (TIGER)}
& Scratch 
& \angcell{25.8}{ 32.5}{ 41.6}{ 50.5}
& \angcell{24.8}{ 27.1}{ 38.5}{ 46.6}
& \angcell{22.5}{ 29.0}{ 39.3}{ 49.1}
& \angcell{19.0}{ 24.8}{ 35.0}{ 44.4}
& 36.3
\\
& PT+FT 
& \angcell{19.9}{ 25.9}{ 39.8}{ 49.1}
& \angcell{19.8}{ 22.2}{ 36.1}{ 45.1}
& \angcell{18.6}{ 24.4}{ 36.3}{ 45.6}
& \angcell{15.7}{ 20.1}{ 30.9}{ 40.9}
& 32.8
\\

\midrule

\multirow{2}{*}{\textbf{PATSE}}
& Scratch 
& \angcell{15.6}{ 19.6}{ 24.7}{ 28.0}
& \angcell{12.9}{ 22.5}{ 22.7}{ 22.5}
& \angcell{10.9}{ 20.0}{ 15.4}{ 19.7}
& \angcell{10.1}{ 12.6}{ 14.4}{ 20.2}
& 18.9
\\
& \textbf{PT+FT} 
& \textbf{\angcell{15.1}{ 15.3}{ 21.4}{ 22.8}}
& \angcell{\textbf{11.0}}{\textbf{ 9.6}}{\textbf{ 13.3}}{ 14.7}
& \textbf{\angcell{8.9}{ 10.0}{ 13.5}{ 15.3}}
& \angcell{\textbf{8.8}}{\textbf{ 9.5}}{\textbf{ 12.2}}{ 14.6}
& \textbf{14.0}
\\

\bottomrule
\end{tabular}
\end{table*}
}
\noindent\textbf{Baselines.}\label{sec:baseline} Representative front-end and pipeline baselines are compared.
\textit{DSB+Gate} applies delay-and-sum beamforming followed by frame-level energy gating.
\textit{FastMNMF}~\cite{sekiguchi2020fast} is a multi-channel nonnegative matrix factorization (MNMF)-based blind source separation method.
\textit{Sortformer+GSS} performs speaker diarization using Sortformer~\cite{park2024sortformer} followed by guided source separation (GSS)~\cite{Raj2023GPUacceleratedGS}.
For CSS-based systems, only the separation backbone is employed, excluding the proposed spatial encoder and conditioning modules (Figure~\ref{fig:framework}). 
FasNet-TAC~\cite{luo2020end} and TIGER extended with MCFF for multi-channel input are adopted as separation backbones, and oracle speaker assignment is provided for all CSS-based methods.
\vspace{0.4\baselineskip}

\noindent\textbf{System Configuration.} Table~\ref{table:compare} summarizes the configurations of all compared systems.
The \textit{DOA} column indicates whether prior speaker DOA information is required.
\textit{Chunk} describes how long-form recordings are segmented: no segmentation (\textit{None}),  where the full recording is processed as a whole; diarization-based segmentation (\textit{SD}); or uniform segmentation using a 4\,s window with 2\,s overlap (\textit{Uniform}).
\textit{Stitch} denotes cross-chunk permutation alignment, which is based on mean squared error (\textit{MSE}) for CSS systems, whereas \textit{Direct} refers to permutation-free concatenation, as adopted in PATSE.
In this setup, CSS is configured to produce three output streams, enabling separation of up to three overlapping speakers.
\textit{Spk.\ ID} specifies how speaker identities are assigned (\textit{SD} or \textit{Oracle}).
In contrast, PATSE produces target-conditioned outputs and does not require an additional speaker assignment step. 
\vspace{0.4\baselineskip}

\noindent\textbf{ASR and Segmentation.} For ASR evaluation, estimated waveforms are segmented into utterances and transcribed using Whisper Large-v3~\cite{radford2022robust}.
For long-form outputs (e.g., \textit{DSB+Gate}, \textit{FastMNMF}, and PATSE after stitching), Silero-VAD~\cite{Silero_VAD} is applied to obtain utterance boundaries, whereas \textit{Sortformer+GSS} directly produces utterance-level outputs. For CSS-based systems, segmentation is derived from oracle speaker assignment, which defines the speaker boundaries.
\vspace{0.4\baselineskip}

\noindent\textbf{Implementation Details.} 
$\mathcal{T}(\cdot)$ is implemented as a two-layer Conv2D with PReLU, 
and $\mathcal{F}(\cdot)$ is a Conv2D with PReLU. 
All separation backbone configurations, including $F$, $T$, $N$, and $K$, follow TIGER-Large\footnote{https://huggingface.co/JusperLee/TIGER-speech}.
In the loss, the scaling factor $\alpha$ is set to 0.005. 
The model is trained with a learning rate of 5e-4.
ASR is evaluated using word error rate (WER), and speaker diarization using diarization error rate (DER).






\section{Results}
\noindent\textbf{LibriReplay-DOA.}
\label{sec:librireplay}
Table~\ref{table:LibriReplay} reports WER on LibriReplay-DOA across four target–interferer angles (15$^\circ$, 45$^\circ$, 90$^\circ$, 120$^\circ$) and four overlap ranges (0--25\%, 25--50\%, 50--75\%, 75--100\%).
In the \textit{Training Strategy} column, \textit{NT} denotes no training; 
\textit{Scratch} denotes training from scratch on the training data in Section~\ref{sec:train_data}; 
and \textit{PT+FT} denotes initialization from a pretrained model (TIGER-Large) followed by finetuning.
DOA serves as an explicit cue to the target speaker in \textit{DSB+Gate} and PATSE, whereas CSS-based methods and \textit{FastMNMF} are provided with oracle target speaker identities for comparison.


PATSE is applied per target DOA to separate all speakers. 
Since \textit{DSB+Gate} cannot handle heavy overlap, it is evaluated only for overlap $<$25\%. 
As expected, smaller target–interferer angles and higher overlap ratios increase task difficulty.
Overall, PATSE consistently achieves the lowest WER across nearly all configurations, demonstrating the benefit of DOA-conditioned target extraction.
The improvement of PATSE arises not only from DOA conditioning, as \textit{DSB+Gate} also exploits DOA cues yet performs substantially worse, but also from the architectural advantages of the proposed framework.
By extracting a single target stream per speaker, PATSE avoids inter-output crosstalk and cross-chunk permutation ambiguity in CSS systems.
Even with oracle speaker assignment—an unrealistic upper bound that entirely eliminates diarization errors—\textit{CSS(TIGER)} remains substantially worse, implying that the gap would be even larger under realistic conditions.
The diarization-based pipeline \textit{Sortformer+GSS} performs particularly poorly under high overlap conditions, as speaker temporal boundary estimation becomes unreliable, causing cascaded diarization–separation pipelines to struggle.
Interestingly, \textit{FastMNMF} shows competitive performance only under extremely high overlap ratios (75–100\%), a scenario that is rarely encountered in real-world meetings.
\vspace{0.4\baselineskip}

{
\begin{table}[t]
\centering
\fontsize{8pt}{9.5pt}\selectfont
\caption{WER (\%)$\downarrow$ and DER (\%)$\downarrow$ on TEIDAN dataset.}
\label{table:real}
\setlength{\tabcolsep}{6pt}
\renewcommand{\arraystretch}{1.05}

\begin{tabular}{lcc}
\toprule
\textbf{Method} & \textbf{WER (\%)} & \textbf{DER (\%)} \\
\midrule
DSB + Gate        & 41.33 & 35.72 \\
FastMNMF          & 26.82 & 28.42 \\
Sortformer+GSS    & 45.03 & 36.15 \\
CSS(TIGER)        & 37.43 & -- \\
\textbf{PATSE}    & \textbf{20.50} & \textbf{13.83} \\
\bottomrule
\end{tabular}

\end{table}
}
\noexpand{\textbf{TEIDAN Dataset.}} Table~\ref{table:real} reports ASR (measured by WER) and speaker diarization (measured by DER) results on the real-world conversational dataset TEIDAN.
The same experimental setup is used, and the best-performing \textit{CSS(TIGER)} setting is selected to represent CSS-based systems.
Specifically, for \textit{Sortformer+GSS}, the DER is computed directly from the diarization outputs of Sortformer.
PATSE achieves the best overall performance in terms of both WER and DER, demonstrating improved robustness for speaker-attributed transcription in real-world three-party conversations.
The presence of filler words and informal expressions can increase recognition difficulty; thus, the WER may be partly limited by the ASR backend.

\section{Conclusion}
In this paper, we presented PATSE, a position-aware target speaker extraction framework for addressing the ``who spoke when and what'' problem in multi-party conversations.
By producing target-conditioned, speaker-attributed streams, PATSE derived speaker activity via simple post-processing without explicit diarization.
To facilitate evaluation of DOA-based methods, we introduced and released LibriReplay-DOA, a real-room dataset with DOA annotations to address the lack of DOA labels in real-room recordings.
Experiments on LibriReplay-DOA and the real-world conversational TEIDAN dataset demonstrated consistent improvements in downstream ASR compared to conventional pipelines, even under oracle speaker assignment.


\section{Acknowledgments}
This work was supported by JST BOOST JPMJBS2407 and JST Moonshot R\&D JPMJMS2011.

\section{Generative AI Use Disclosure}
Generative AI tools (Gemini and ChatGPT) were used for language editing and improving the phrasing of this manuscript.

\bibliographystyle{IEEEtran}
\bibliography{mybib}

@inproceedings{cornell2024one,
  title={One model to rule them all? towards end-to-end joint speaker diarization and speech recognition},
  author={Cornell, Samuele and Jung, Jee-weon and Watanabe, Shinji and Squartini, Stefano},
  booktitle={Proc. ICASSP},
  pages={11856--11860},
  year={2024},
}

@article{diaz2021gpurir,
  title={{gpuRIR}: A python library for room impulse response simulation with {GPU} acceleration},
  author={Diaz-Guerra, David and Miguel, Antonio and Beltran, Jose R},
  journal={Multimedia Tools and Applications},
  volume={80},
  number={4},
  pages={5653--5671},
  year={2021},
  publisher={Springer}
}

@article{niu2025dcf,
  title={{DCF-DS}: Deep cascade fusion of diarization and separation for speech recognition under realistic single-channel conditions},
  author={Niu, Shu-Tong and Du, Jun and Wang, Ruo-Yu and Yang, Gao-Bin and Gao, Tian and Pan, Jia and Hu, Yu},
  journal={IEEE/ACM Trans. ASLP},
  year={2025},
  publisher={IEEE}
}

@inproceedings{chen2020continuous,
  title={Continuous speech separation: Dataset and analysis},
  author={Chen, Zhuo and Yoshioka, Takuya and Lu, Liang and Zhou, Tianyan and Meng, Zhong and Luo, Yi and Wu, Jian and Xiao, Xiong and Li, Jinyu},
  booktitle={Proc. ICASSP},
  pages={7284--7288},
  year={2020},
}

@inproceedings{raj2021integration,
  title={Integration of speech separation, diarization, and recognition for multi-speaker meetings: System description, comparison, and analysis},
  author={Raj, Desh and Denisov, Pavel and Chen, Zhuo and Erdogan, Hakan and Huang, Zili and He, Maokui and Watanabe, Shinji and Du, Jun and Yoshioka, Takuya and Luo, Yi and others},
  booktitle={Proc. SLT},
  pages={897--904},
  year={2021},
}

@inproceedings{taherian2023multi,
  title={Multi-resolution location-based training for multi-channel continuous speech separation},
  author={Taherian, Hassan and Wang, DeLiang},
  booktitle={Proc. ICASSP},
  pages={1--5},
  year={2023},
}

@inproceedings{yoshioka2019low,
  title={Low-latency speaker-independent continuous speech separation},
  author={Yoshioka, Takuya and Chen, Zhuo and Liu, Changliang and Xiao, Xiong and Erdogan, Hakan and Dimitriadis, Dimitrios},
  booktitle={Proc. ICASSP},
  pages={6980--6984},
  year={2019},
}

@inproceedings{von2021graph,
  title={Graph-{PIT}: Generalized Permutation Invariant Training for Continuous Separation of Arbitrary Numbers of Speakers},
  author={von Neumann, Thilo and Kinoshita, Keisuke and Boeddeker, Christoph and Delcroix, Marc and Haeb-Umbach, Reinhold},
  booktitle={Proc. Interspeech},
  pages={3490--3494},
  year={2021}
}

@inproceedings{delcroix2021speaker,
  title={Speaker activity driven neural speech extraction},
  author={Delcroix, Marc and Zmolikova, Katerina and Ochiai, Tsubasa and Kinoshita, Keisuke and Nakatani, Tomohiro},
  booktitle={Proc. ICASSP},
  pages={6099--6103},
  year={2021},
}

@inproceedings{raj2022gpu,
  title={GPU-accelerated Guided Source Separation for Meeting Transcription},
  author={Raj, Desh and Povey, Daniel and Khudanpur, Sanjeev},
  booktitle={Proc. Interspeech},
  pages={3507--3511},
  year={2023}
}

@inproceedings{medennikov2020stc,
  title={The {STC} system for the CHiME-6 challenge},
  author={Medennikov, Ivan and Korenevsky, Maxim and Prisyach, Tatiana and Khokhlov, Yuri and Korenevskaya, Mariya and Sorokin, Ivan and Timofeeva, Tatiana and Mitrofanov, Anton and Andrusenko, Andrei and Podluzhny, Ivan and others},
  booktitle={CHiME 2020 Workshop on Speech Processing in Everyday Environments},
  year={2020}
}

@inproceedings{fujita2019end,
  title={End-to-End Neural Speaker Diarization with Permutation-Free Objectives},
  author={Fujita, Yusuke and Kanda, Naoyuki and Horiguchi, Shota and Nagamatsu, Kenji and Watanabe, Shinji},
  booktitle={Proc. Interspeech},
  pages={4300--4304},
  year={2019}
}

@article{taherian2024multi,
  title={Multi-channel conversational speaker separation via neural diarization},
  author={Taherian, Hassan and Wang, DeLiang},
  journal={IEEE/ACM Trans. ASLP},
  volume={32},
  pages={2467--2476},
  year={2024},
  publisher={IEEE}
}

@inproceedings{wang2024study,
  title={A study of multichannel spatiotemporal features and knowledge distillation on robust target speaker extraction},
  author={Wang, Yichi and Zhang, Jie and Chen, Shihao and Zhang, Weitai and Ye, Zhongyi and Zhou, Xinyuan and Dai, Lirong},
  booktitle={Proc. ICASSP},
  pages={431--435},
  year={2024},
}

@inproceedings{wang2025leveraging,
  title={Leveraging Boolean Directivity Embedding for Binaural Target Speaker Extraction},
  author={Wang, Yichi and Zhang, Jie and Jiang, Chengqian and Zhang, Weitai and Ye, Zhongyi and Dai, Lirong},
  booktitle={Proc. ICASSP},
  pages={1--5},
  year={2025},
}

@inproceedings{briegleb2023exploiting,
  title={Exploiting spatial information with the informed complex-valued spatial autoencoder for target speaker extraction},
  author={Briegleb, Annika and Halimeh, Mhd Modar and Kellermann, Walter},
  booktitle={Proc. ICASSP},
  pages={1--5},
  year={2023},
}

@inproceedings{shi2026train,
  title={Train short, infer long: Speech-llm enables zero-shot streamable joint asr and diarization on long audio},
  author={Shi, Mohan and Xiao, Xiong and Fan, Ruchao and Ling, Shaoshi and Li, Jinyu},
  booktitle={Proc. ICASSP},
  pages={17442--17446},
  year={2026},
}

@inproceedings{shi2023casa,
  title={CASA-ASR: Context-Aware Speaker-Attributed ASR},
  author={Shi, Mohan and Du, Zhihao and Chen, Qian and Yu, Fan and Li, Yangze and Zhang, Shiliang and Zhang, Jie and Dai, Li-Rong},
  booktitle={Proc. Interspeech},
  pages={411--415},
  year={2023}
}

@inproceedings{elminshawi2023beamformer,
  title={Beamformer-guided target speaker extraction},
  author={Elminshawi, Mohamed and Chetupalli, Srikanth Raj and Habets, Emanu{\"e}l AP},
  booktitle={Proc. ICASSP},
  pages={1--5},
  year={2023},
}

@inproceedings{wang2018voicefilter,
  title={Voice{F}ilter: Targeted Voice Separation by Speaker-Conditioned Spectrogram Masking},
  author={Wang, Quan and Muckenhirn, Hannah and Wilson, Kevin and Sridhar, Prashant and Wu, Zelin and Hershey, John R and Saurous, Rif A and Weiss, Ron J and Jia, Ye and Moreno, Ignacio Lopez},
  booktitle={Proc. Interspeech},
  pages={2728--2732},
  year={2019}
}

@inproceedings{delcroix2019end,
  title={End-to-{E}nd SpeakerBeam for Single Channel Target Speech Recognition.},
  author={Delcroix, Marc and Watanabe, Shinji and Ochiai, Tsubasa and Kinoshita, Keisuke and Karita, Shigeki and Ogawa, Atsunori and Nakatani, Tomohiro},
  booktitle={Proc. Interspeech},
  pages={451--455},
  year={2019}
}

@article{boeddeker2024ts,
  title={{TS-SEP}: Joint diarization and separation conditioned on estimated speaker embeddings},
  author={Boeddeker, Christoph and Subramanian, Aswin Shanmugam and Wichern, Gordon and Haeb-Umbach, Reinhold and Le Roux, Jonathan},
  journal={IEEE/ACM Trans. ASLP},
  volume={32},
  pages={1185--1197},
  year={2024},
  publisher={IEEE}
}

@article{zhang2025doa,
  title={{DOA} or {S}peaker Embedding: Which is better for multi-microphone target speaker extraction},
  author={Zhang, Shuang and Zhang, Jie and Wang, Yichi and Yan, Haoyin},
  journal={IEEE Signal Processing Letters},
  year={2025},
  publisher={IEEE}
}

@inproceedings{elmers2025triadic,
  title={Triadic Multi-party Voice Activity Projection for Turn-taking in Spoken Dialogue Systems},
  author={Elmers, Mikey and Inoue, Koji and Lala, Divesh and Kawahara, Tatsuya},
  booktitle={Proc. Interspeech},
  pages={3015--3019},
  year={2025}
}

@article{xu2024tiger,
  title={Tiger: Time-frequency interleaved gain extraction and reconstruction for efficient speech separation},
  author={Xu, Mohan and Li, Kai and Chen, Guo and Hu, Xiaolin},
  journal={arXiv preprint arXiv:2410.01469},
  year={2024}
}

@inproceedings{luo2020end,
  title={End-to-end microphone permutation and number invariant multi-channel speech separation},
  author={Luo, Yi and Chen, Zhuo and Mesgarani, Nima and Yoshioka, Takuya},
  booktitle={Proc. ICASSP},
  pages={6394--6398},
  year={2020},
}

@inproceedings{perez2018film,
  title={Film: Visual reasoning with a general conditioning layer},
  author={Perez, Ethan and Strub, Florian and De Vries, Harm and Dumoulin, Vincent and Courville, Aaron},
  booktitle={Proc. AAAI.},
  volume={32},
  number={1},
  year={2018}
}

@article{pan2022usev,
  title={{USEV}: Universal speaker extraction with visual cue},
  author={Pan, Zexu and Ge, Meng and Li, Haizhou},
  journal={IEEE/ACM Trans. ASLP},
  volume={30},
  pages={3032--3045},
  year={2022},
  publisher={IEEE}
}

@inproceedings{panayotov2015librispeech,
  title={Librispeech: an asr corpus based on public domain audio books},
  author={Panayotov, Vassil and Chen, Guoguo and Povey, Daniel and Khudanpur, Sanjeev},
  booktitle={Proc. ICASSP},
  pages={5206--5210},
  year={2015},
}

@article{reddy2020interspeech,
  title={The interspeech 2020 deep noise suppression challenge: Datasets, subjective testing framework, and challenge results},
  author={Reddy, Chandan KA and Gopal, Vishak and Cutler, Ross and Beyrami, Ebrahim and Cheng, Roger and Dubey, Harishchandra and Matusevych, Sergiy and Aichner, Robert and Aazami, Ashkan and Braun, Sebastian and others},
  journal={arXiv preprint arXiv:2005.13981},
  year={2020}
}

@article{park2024sortformer,
  title={Sortformer: A novel approach for permutation-resolved speaker supervision in speech-to-text systems},
  author={Park, Taejin and Medennikov, Ivan and Dhawan, Kunal and Wang, Weiqing and Huang, He and Koluguri, Nithin Rao and Puvvada, Krishna C and Balam, Jagadeesh and Ginsburg, Boris},
  journal={arXiv preprint arXiv:2409.06656},
  year={2024}
}

@article{Raj2023GPUacceleratedGS,
  title={{GPU}-accelerated guided source separation for meeting transcription},
  author={Raj, Desh and Povey, Daniel and Khudanpur, Sanjeev},
  journal={arXiv preprint arXiv:2212.05271},
  year={2022}
}

@article{sekiguchi2020fast,
  title={Fast multichannel nonnegative matrix factorization with directivity-aware jointly-diagonalizable spatial covariance matrices for blind source separation},
  author={Sekiguchi, Kouhei and Bando, Yoshiaki and Nugraha, Aditya Arie and Yoshii, Kazuyoshi and Kawahara, Tatsuya},
  journal={IEEE/ACM Trans. ASLP},
  volume={28},
  pages={2610--2625},
  year={2020},
  publisher={IEEE}
}

@misc{Silero_VAD,
  author = {Silero Team},
  title = {Silero {VAD}: pre-trained enterprise-grade Voice Activity Detector ({VAD}), Number Detector and Language Classifier},
  year = {2024},
  publisher = {GitHub},
  journal = {GitHub repository},
  howpublished = {\url{https://github.com/snakers4/silero-vad}},
  commit = {insert_some_commit_here},
  email = {hello@silero.ai}
}

@article{radford2022robust,
  title={Robust Speech Recognition via Large-Scale Weak Supervision},
  author={Radford, Alec and others},
  year={2022},
  eprint={2212.04356},
  archivePrefix={arXiv},
  primaryClass={cs.CL}
}

@article{wang2018combining,
  title={Combining spectral and spatial features for deep learning based blind speaker separation},
  author={Wang, Zhong-Qiu and Wang, DeLiang},
  journal={IEEE/ACM Trans. ASLP},
  volume={27},
  number={2},
  pages={457--468},
  year={2018},
  publisher={IEEE}
}

\end{document}